# ON PROPERTIES OF THE ISOSCALAR GIANT DIPOLE RESONANCE

M.L. Gorelik, M.H. Urin [*]

*Moscow State Engineering Physics Institute (Technical University), 115409 Moscow, Russia*

## Abstract

Main properties (strength function, energy-dependent transition density, branching ratios for direct nucleon decay) of the isoscalar giant dipole resonance in several medium-heavy mass spherical nuclei are described within a continuum-RPA approach, taking into account the smearing effect. All model parameters used in the calculations are taken from independent data. Calculation results are compared with available experimental data.

1. Recently several experimental [1,2] and theoretical [3-6] works have been published describing results of studies of properties of the isoscalar giant dipole resonance (ISGDR) in several medium-heavy mass spherical nuclei. It has been found in Ref. [2], from an analysis of $(\alpha, \alpha')$-reaction at small angles, that the isoscalar dipole strength distribution exhibits two main regions of strength concentration, corresponding to the lower (pygmy) and upper (main) ISGDR components. Microscopic approaches used in recent theoretical studies of the ISGDR are based on: (i) continuum-RPA calculations with the use of the Landau-Migdal particle-hole interaction [3], (ii) Hartree-Fock+RPA calculations with the use of the Skyrme interactions [4,6], and (iii) relativistic RPA calculations [5]. In each of these approaches, the strength distribution of the ISGDR shows two main regions of strength concentration with corresponding centroid energies which are in qualitative agreement with those of Ref. [2]. References to previous experimental and theoretical studies of the ISGDR are given, respectively, in Refs. [1,2] and [3-6]. Here, we mention Ref. [7], where the low-energy isoscalar $1^-$ strength has been identified from an analysis of $(\alpha\alpha', \gamma)$-reaction.

---

[*]E-mail: urin@theor.mephi.ru

In connection with the above-mentioned investigations it seems reasonable to realize the next step in theoretical studies, which consists in a rather full description of ISGDR properties. Such a description includes the calculations of: (i) the ISGDR strength distribution in a wide interval of excitation-energy, taking into account the smearing effect, (ii) the energy-dependent ISGDR transition density also in a wide energy interval, and (iii) the partial branching ratios for direct nucleon decay of the ISGDR. In each of the above-mentioned theoretical approaches, used in earlier works, this program was only partially realized. In the present work we attempt to describe the above-listed ISGDR characteristics in an extended version of the continuum-RPA (CRPA) approach of Ref. [3]. Calculation results obtained for $^{90}$Zr, $^{116}$Sn, $^{144}$Sm and $^{208}$Pb are compared with available experimental data.

2. Apart from the description of some ISGDR properties, the partially self-consistent continuum-RPA approach was mainly used in Ref. [3] to describe in a quantitative way the direct neutron decay of the isoscalar giant monopole resonance (ISGMR). To realize a rather full description of ISGDR properties, we extend the approach of Ref. [3] in the following ways. (For brevity, we use below the notations of Ref. [3] and sometimes refer to equations from this Ref.) First, we slightly change the dimensionless parameters $f^{in}$ and $f^{ex}$ of the radial-dependent intensity $F(r)$ of the isoscalar part of the Landau-Migdal particle-hole interaction (determined by Eq.(16) of Ref. [3]) to better describe the experimental energies of the ISGMR (taken from Ref. [8]). The new value $f^{in} = 0.0875$ (as well as the value $f^{in} = -0.0875$ used in Ref. [3]) is in agreement with the systematics of the Landau-Migdal parameters of Ref. [9]. As in Ref. [3], the $f^{ex}$ value is adjusted to make the $1^-$ spurious-state energy close to zero for each considered nucleus (see Table 1). The relative strengths $x_{s.s.}$ of the spurious state (s.s.), or the percentages of the respective energy-weighted sum rule (EWSR) exhausted by the s.s. (Eq.(18) of Ref. [3]), are also given in Table 1.

To calculate the energy-averaged strength functions of the ISGMR (L=0) and the ISGDR (L=1) $\bar{S}_L(\omega)$, accounting for the smearing effect, we solve the CRPA equations of Ref. [3] with the replacement of the excitation energy $\omega$ by $\omega + \frac{i}{2}I(\omega)$: $\bar{S}_L(\omega) = S_L(\omega + \frac{i}{2}I(\omega))$. The smearing parameter $I(\omega)$ (the mean doorway-state spreading width) is taken from Ref. [10] with the energy-dependent function having a saturation-like behaviour. A reasonable description of the total width has been obtained in Ref. [11] for several isovector giant resonances with the use of the $I(\omega)$ from Ref. [10]. The relative energy-weighted strength functions $y_L(\omega) = \omega \bar{S}_L(\omega)/(EWSR)_L$ calculated for the ISGMR and ISGDR allow us to deduce for some excitation-energy intervals $\omega_1 - \omega_2$ the following parameters: centroid of the energy $\omega_L$, root mean square (RMS) width $\Delta_L$, relative strength $x_L$. These parameters are shown in Table 1 for the ISGMR and in Table 2 for the ISGDR. Some of these results are compared with available experimental data in Table 3. The calculated strength functions $y_{L=1}(\omega)$ are shown in Fig.1.

The giant-resonance transition density $\rho_L(r)$ can reasonably be defined in the CRPA in the special case, when only one collective particle-hole-type state (doorway state) corresponds to the considered GR and, therefore, exhausts most of the respective $EWSR$. Such a situation takes place in the approach of Ref. [3] for the ISGMR. In the case of the ISGDR, several doorway states have comparable strength (see e.g. Ref. [3]) and, therefore, only the energy-averaged and energy-dependent transition density $\bar{\rho}_L(r,\omega)$ can be defined. In accordance with the spectral expansion for the effective particle-hole propagator (the particle-hole Green's function) one can get the expression:

$$\bar{\rho}_L(r,\omega) = -\frac{1}{\pi} \frac{\operatorname{Im} \sum_{\alpha=n,p} \tilde{V}_{L,\alpha}(r, \omega + \frac{i}{2}I(\omega))}{2F(r)\bar{S}_L^{1/2}(\omega)}, \quad (1)$$

which is equivalent to that used in Ref. [6]. In Eq.(1), $\tilde{V}_{L,\alpha}(r,\omega)$ are the effective fields (defined by Eq.(2) of Ref. [3]) corresponding to the probe operator $V_L(r)$ : $V_{L=0} = r^2, V_{L=1} = r^3 - \eta r$ with $\eta = \frac{5}{3} < r^2 >$ [3,4,6]. From Eq.(1) and Eqs.(1) and (2) of Ref. [3] follows the expression $\bar{S}_L(\omega) = (\int V_L(r)\bar{\rho}_L(r,\omega)r^2 dr)^2$, which is in agreement with the definitions of Ref. [6]. As applied to $^{208}$Pb, the properly normalized transition density $r^2 \bar{\rho}_{L=1}(r,\omega)/\bar{S}_{L=1}^{1/2}(\omega)$ is shown in Fig.2 for some of values of $\omega$ in a comparison with the collective ISGDR transition density [12] normalized by the same way.

To calculate the partial and total branching ratios for direct nucleon decay of the main ISGDR component, we follow Refs. [11] and [13], where the proton branching ratios have been estimated for the high-energy charge-exchange spin-monopole and monopole giant resonances, respectively:

$$b_{\mu,\alpha} = \frac{\sum_{(\lambda)} \int_{\omega_1}^{\omega_2} |\bar{M}_c^L(\omega)|^2 d\omega}{\int_{\omega_1}^{\omega_2} \bar{S}_L(\omega) d\omega}; \quad b_\alpha = \sum_\mu b_{\mu,\alpha}. \qquad (2)$$

Here, $\bar{M}_c^L(\omega) = M_c^L(\omega + \frac{i}{2}I(\omega))$ is the energy-averaged reaction amplitude corresponding to direct nucleon decay with population of one-hole state $\mu^{-1}$ in the product nucleus; $c = \mu, \alpha, (\lambda), \varepsilon$ is the set of decay-channel quantum numbers, which includes the energy $\varepsilon = \omega + \varepsilon_\mu$ and quantum numbers $(\lambda) = j, l$ of the escaped nucleon. The definition of CRPA reaction-amplitude $M_c^L(\omega)$ is given by Eq.(5) of Ref. [3]. Note, that in the CRPA ($I = 0$) the total branching ratio $b = b_n + b_p$ is equal unity by definition. The partial branching ratios $b_{\mu,\alpha}$ calculated for the upper component of the ISGDR (15-30 MeV) in $^{208}$Pb are given in Table 4.

3. We now make several comments on the results of this work: (i) With the choice of the Landau-Migdal parameters $f^{in} = 0.0875$ and $f^{ex}$ from Table 1, it is possible within the present CRPA approach to describe satisfactorily the experimental centroids of the energy for the ISGMR and both ISGDR components (Table 1-3). After taking the results of Ref. [7] into account [2] (not shown in Table 3) the theoretical description of the experimental centroid energies of the lower ISGDR improves. (ii) The use of the saturation-like dependence for $I(\omega)$ with parameters taken from independent data [10,11] allows us to describe reasonably the experimental RMS widths for both ISGDR components (Table 3). A similar conclusion for the ISGMR follows from visual comparison of the calculated strength functions $y_{L=0}(\omega)$ (not shown here) with those deduced from experimental data in Ref. [8]. (iii) The calculated relative strength of both ISGDR components (Table 2) are markedly less than the corresponding values deduced from experimental data [2]. Possible reasons for the difference are the use of the specific collective ISGDR transition density for the analysis of the data in Ref. [2], and (or) the neglect in our calculation of the contribution of momentum-dependent forces to the Landau-Migdal particle-hole interaction. Isovector momentum-dependent forces were taken into account in Ref. [10] to describe in the same CRPA approach the main properties of the isovector giant dipole resonance, using the relative effective nucleon mass of unity. For this reason, in the present work unit relative effective mass is also used and, therefore, the isoscalar part of momentum-dependent forces is not considered. As is shown in Table 2, the calculated ISGDR parameters discussed above are markedly dependent on the considered excitation-energy interval. Such a dependence is a result of both the Landau damping and the smearing effect. (iv) The radial dependence of the transition density of the main ISGDR component calculated for $^{208}$Pb is rather close to that found in the scaling model [12] (Fig.2). However, it is not true for the lower component (Fig.2). Thus, the use of the microscopic energy-dependent transition density of Eq.(1) for analyzing experimental cross sections seems preferable. Such an attempt has been recently realized in Ref. [6]. (v) The calculated branching ratios for direct nucleon decay of the upper ISGDR component are rather large due to a strong coupling of this component to the continuum. The difference between the present calculated results for the branching ratios and the previous results of Ref. [3] is partially explained by a large contribution (due to the smearing effect) of the "tail" of the ISGDR lower component to the energy-averaged reaction amplitudes.

In conclusion, we have described the main properties of the ISGDR in several medium-heavy mass spherical nuclei using a transparent and rather easy to implement approach which is based on continuum-RPA method with the inclusion of the smearing effect. Except for the relative strengths, a satisfactory description of available experimental data on parameters of the ISGDR components was obtained. Following Ref. [6], we suggest the use of the microscopic energy-dependent transition density of the ISGDR for the analysis of experimental cross sections. Such use of the transition density allows one to clarify the problem with the underestimation of the calculated ISGDR relative strength in comparison to that deduced from experimental data.

The authors are gratefull to S. Shlomo for interesting discussions and valuable remarks.

**Figure 1:** The calculated relative strength function $y_{L=1}(\omega)$. The dash-dotted, full, dashed and thin lines are for $^{90}$Zr, $^{116}$Sn, $^{144}$Sm and $^{208}$Pb, respectively.

**Figure 2:** The normalized ISGDR transition density (in arbitrary units) calculated at several energies (the full, dotted, dash-dotted and dashed lines correspond to 23.06, 11.26, 7.76, and 6.81 MeV, respectively) in comparison with the normalized transition density (thin line) calculated in the scaling model [12].

TABLES

TABLE I. The relative isoscalar dipole strength of the spurious state and parameters of the ISGMR(L=0) calculated with the use of shown Landau-Migdal parameters $f^{ex}$ and of value $f^{in} = 0.0875$.

| A | $-f^{ex}$ | $x_{s.s.}$ [%] | $\omega_1 - \omega_2$ [MeV] | $\omega_L$ [MeV] | $\Delta_L$ [MeV] | $x_L$ [%] |
|---|---|---|---|---|---|---|
| $^{208}Pb$ | 2.897 | 91.9 | 10-20 | 14.29 | 2.05 | 80.2 |
|  |  |  | 3-60 | 15.22 | 5.13 | 99.2 |
| (a) |  |  | 10-20 | 13.99 | 1.10 | 97.9 |
| $^{144}Sm$ | 2.811 | 93.6 | 10-20 | 15.28 | 2.00 | 78.0 |
|  |  |  | 3-60 | 16.53 | 5.16 | 98.5 |
| (a) |  |  | 10-20 | 15.27 | 0.96 | 98.7 |
| $^{116}Sn$ | 2.832 | 93.6 | 10-20 | 15.79 | 2.08 | 74.7 |
|  |  |  | 3-60 | 17.18 | 5.29 | 98.4 |
| (a) |  |  | 10-20 | 15.97 | 1.24 | 97.9 |
| $^{90}Zr$ | 2.753 | 94.5 | 10-25 | 17.10 | 2.71 | 85.4 |
|  |  |  | 3-60 | 18.05 | 5.30 | 98.2 |
| (a) |  |  | 10-25 | 16.89 | 1.35 | 99.8 |

(a) I=0.05 MeV.

TABLE II. Parameters of the ISGDR(L=1) calculated for different excitation-energy intervals.

| A | Lower ISGDR | | | | Upper ISGDR | | | |
|---|---|---|---|---|---|---|---|---|
| | $\omega_1 - \omega_2$ [MeV] | $\omega_L$ [MeV] | $\Delta_L$ [MeV] | $x_L$ [%] | $\omega_1 - \omega_2$ [MeV] | $\omega_L$ [MeV] | $\Delta_L$ [MeV] | $x_L$ [%] |
| $^{208}Pb$ | 8-15 | 11.10 | 1.91 | 13.3 | 15-24 | 20.71 | 2.41 | 41.3 |
| | 5-15 | 9.87 | 2.52 | 16.7 | 15-30 | 22.57 | 3.36 | 68.6 |
| | | | | | 15-60 | 24.03 | 5.83 | 81.1 |
| (a) | 5-15 | 9.67 | 2.30 | 18.2 | 15-30 | 22.75 | 2.49 | 79.7 |
| $^{144}Sm$ | 5-15 | 10.74 | 2.19 | 12.3 | 15-35 | 24.38 | 4.13 | 76.4 |
| | | | | | 15-60 | 25.43 | 6.00 | 84.8 |
| (a) | 5-15 | 10.64 | 1.84 | 13.9 | 15-30 | 24.18 | 2.75 | 81.6 |
| $^{116}Sn$ | 11-18 | 14.02 | 2.01 | 10.8 | 18-32 | 25.21 | 3.33 | 65.5 |
| | 5-15 | 10.36 | 2.42 | 13.2 | 15-35 | 24.90 | 4.38 | 74.7 |
| | | | | | 15-60 | 26.10 | 6.28 | 84.3 |
| (a) | 5-15 | 10.31 | 2.25 | 15.0 | 15-35 | 25.09 | 3.42 | 83.3 |
| $^{90}Zr$ | 11-18 | 13.89 | 2.08 | 9.9 | 18-32 | 25.64 | 3.52 | 64.5 |
| | 5-16 | 11.42 | 2.23 | 11.3 | 16-40 | 26.30 | 4.93 | 79.7 |
| | | | | | 16-60 | 27.13 | 6.37 | 85.7 |
| (a) | 5-16 | 11.19 | 1.70 | 12.3 | 16-40 | 26.10 | 3.93 | 87.3 |

(a) I=0.05 MeV.

TABLE III. Comparison of parameters calculated for the ISGMR and ISGDR with the corresponding experimental data taken from Ref.[8] and [2], respectively. All the parameters are given in MeV.

| A | ISGMR $\omega_L$ | | Lower ISGDR $\omega_L$ | | $\Delta_L$ | | Upper ISGDR $\omega_L$ | | $\Delta_L$ | |
|---|---|---|---|---|---|---|---|---|---|---|
| $^{208}Pb$ | 14.17±0.28 | 14.3 | 12.2±0.6 | 11.1 | 1.9±0.5 | 1.9 | 19.9±0.8 | 20.7 | 2.5±0.6 | 2.4 |
| $^{144}Sm$ | 15.39±0.28 | 15.3 | | | | | | | | |
| $^{116}Sn$ | 16.07±0.12 | 15.8 | 14.7±0.5 | 14.0 | 1.6±0.5 | 2.0 | 23.0±0.6 | 25.2 | 3.7±0.5 | 3.3 |
| $^{90}Zr$ | 17.89±0.20 | 17.1 | 16.2±0.8 | 13.9 | 1.9±0.7 | 2.1 | 25.7±0.7 | 25.6 | 3.5±0.6 | 3.5 |

TABLE IV. Calculated partial branching ratios for direct nucleon decay of the ISGDR in $^{208}$Pb. The results for decays with population of one-hole states from the last filled shells are shown for excitation-energy interval 15-30 MeV. Spectroscopic factors of these states $S_\mu = 1$ are taken for all decay-channels.

| neutron, $\mu^{-1}$ | $(1/2)^-$ | $(5/2)^-$ | $(3/2)^-$ | $(13/2)^+$ | $(7/2)^-$ | $(9/2)^-$ |
|---|---|---|---|---|---|---|
| $b_\mu$, [%] | 1.4 | 4.8 | 3.4 | 8.0 | 8.5 | 3.9 |
| proton, $\mu^{-1}$ | $(1/2)^+$ | $(3/2)^+$ | $(11/2)^-$ | $(5/2)^+$ | $(7/2)^+$ | |
| $b_\mu$, [%] | 3.0 | 3.9 | 2.4 | 6.1 | 1.4 | |

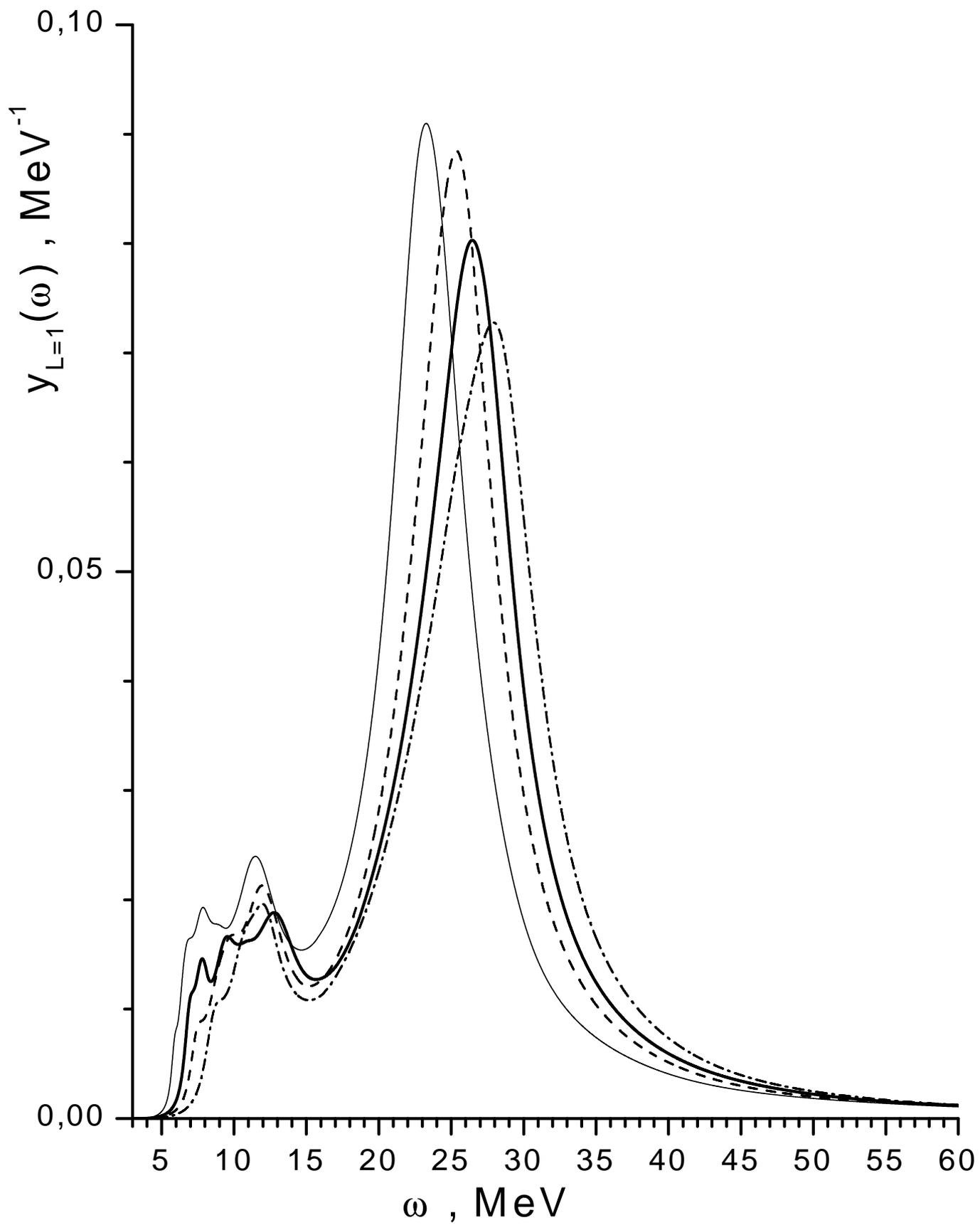

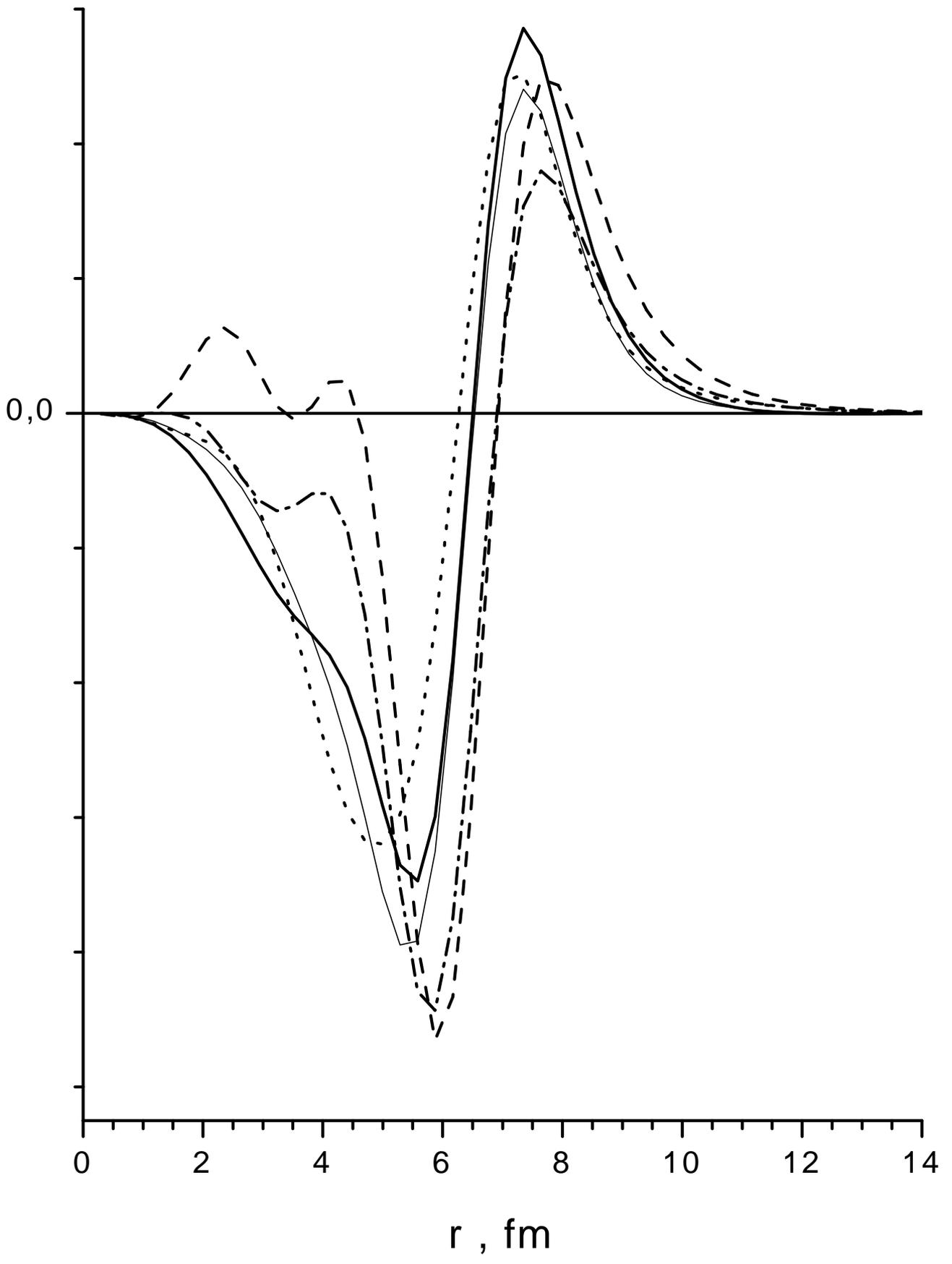